\documentclass[aps,floatfix,nofootinbib,showpacs,twocolumn]{revtex4}

\usepackage{graphicx}
\usepackage{dcolumn}
\usepackage{amsmath}

\begin{document}


\title{%
\hspace*{\fill}{\tt\normalsize ANL-PHY-12078-TH-2008, UMSNH-IFM-F-2008-20}\\[1ex]
Regarding confinement and dynamical chiral symmetry breaking in QED3}

\author{A.~Bashir}
\affiliation{Instituto de F\'{\i}sica y Matem\'aticas, Universidad Michoacana de San Nicol\'as de Hidalgo, Apartado Postal 2-82, Morelia, Michoac\'an 58040, M\'exico}

\author{I.\,C.~Clo\"et}
\affiliation{Physics Division, Argonne National Laboratory,
             Argonne, IL 60439-4843 U.S.A.}

\author{A.~Raya}
\affiliation{Instituto de F\'{\i}sica y Matem\'aticas, Universidad Michoacana de San Nicol\'as de Hidalgo, Apartado Postal 2-82, Morelia, Michoac\'an 58040, M\'exico}

\author{C.\,D.~Roberts}
\affiliation{Physics Division, Argonne National Laboratory,
             Argonne, IL 60439-4843 U.S.A.}

\begin{abstract}
\rule{0ex}{3ex}
We establish that QED3 can possess a critical number of flavours, $N_f^c$, associated with dynamical chiral symmetry breaking if, and only if, the fermion wave function renormalisation and photon vacuum polarisation are homogeneous functions at infrared momenta when the fermion mass function vanishes.  The Ward identity entails that the fermion-photon vertex possesses the same property and ensures a simple relationship between the homogeneity degrees of each of these functions.  Simple models for the photon vacuum polarisation and fermion-photon vertex are used to illustrate these observations.  The existence and value of $N_f^c$ are contingent upon the precise form of the vertex but any discussion of gauge dependence is moot.  We introduce an order parameter for confinement.  Chiral symmetry restoration and deconfinement are coincident owing to an abrupt change in the analytic properties of the fermion propagator when a nonzero scalar self-energy becomes insupportable.
\end{abstract}

\maketitle

\section{Introduction}
\label{sec:one}
Analogous to quenched quantum chromodynamics (QCD), quenched quantum electrodynamics in three dimensions (2 spatial and 1 temporal - QED3) is confining because it has a nonzero string tension \cite{Gopfert:1981er}.  This feature persists in the unquenched theory if massive fermions circulate in the photon vacuum polarisation \cite{Burden:1991uh}.  That mass can either be explicit or dynamical in origin.  The possibility of dynamical mass generation in the chiral limit theory is of material interest because then QED3 has even greater qualitative similarity with QCD.

Dynamical chiral symmetry breaking (DCSB) explains the origin of constituent-quark masses in QCD and underlies the success of chiral effective field theory \cite{Roberts:2007ji}.  Its importance in QCD has ensured that the possibility of DCSB in QED3 has received much attention.  The Dyson-Schwinger equations (DSEs) provide a natural framework within which to explore this and related phenomena.  These methods are reviewed in Refs.\,\cite{Roberts:1994dr,Roberts:2000aa,Alkofer:2000wg,Maris:2003vk,Roberts:2007jh}.

Interest was stimulated by a finding \cite{Appelquist:1988sr} that in QED3 with $N_f$ massless fermions a truncated system of DSEs yields a critical number of flavours, $N_f^c$, such that DCSB is only possible for $N_f<N_f^c$.  A debate has subsequently ensued, which focuses on the reliability of the truncations employed and offers an alternative possibility; namely, that the mass function does not vanish at some critical number of flavours but is instead exponentially suppressed with increasing $N_f$ \cite{Pennington:1990bx}.  Our analysis is intended to crystallise the issues and thereby eliminate the points of contention.

It is worth remarking that lattice-regularised QED3 can currently state only that $N_f^c>1$, if it exists at all \cite{Hands:2004bh}.  An impediment to reliable lattice calculations is the mass hierarchy feature of QED3; viz., any dynamically generated mass-scale is at least one order of magnitude smaller than the natural scale, which is set by the dimensioned coupling $e^2$ (see, e.g., Ref.\,\cite{Bashir:2005wt}).\footnote{This is fortunately not the case in QCD, which possesses a dimensionless but running coupling.  Its evolution is determined by a mass-scale whose magnitude is characteristic of all dynamically generated mass-dimensioned quantities in the theory.}

In parallel with its relevance as a tool through which to develop insight into aspects of QCD, QED3 is also of interest in condensed matter physics as an effective field theory for high-temperature superconductors \cite{Franz:2002qy,Herbut:2002yq,Thomas:2006bj} and graphene \cite{Novoselov:2005kj,Gusynin:2007ix}.  Naturally, the nature of the theory at $N_f=2$ is most interesting for true electronic systems.

We describe some basic features of QED3 in Sec.\,\ref{sec:coupled} and introduce the coupled DSEs relevant to our analysis.  Section~\ref{theory} canvasses the issue of gauge covariance in truncations of the DSEs.  It also elucidates necessary and sufficient conditions for the existence of a critical number of flavours, in particular a collusion between Schwinger functions at infrared momenta.  This is illustrated using \emph{Ans\"atze} for the vacuum polarisation and fermion-photon vertex in Sec.\,\ref{sec:illustration}.  In addition, we introduce a model-independent order parameter for deconfinement and demonstrate simultaneity of chiral symmetry restoration and deconfinement.  We expect that this result remains valid with all consistent truncations.  Section~\ref{apercu} provides a summary and perspective.

\section{Coupled Equations}
\label{sec:coupled}
In three dimensions there are two inequivalent $2\times 2$ representations of the Euclidean Clifford algebra $\{\gamma_\mu,\gamma_\nu\} = 2\delta_{\mu\nu}$.  Hence, two-component spinors are sufficient to describe spinorial representations of the Lorentz Group.  In this case, however, a mass term of any origin is odd under parity transformations.  That can be avoided by employing four-component spinors and a $4\times 4$ representation of the Clifford algebra e.g., the set $\{\gamma_1,\gamma_2,\gamma_4\}$, taken from four dimensional theories, with $\gamma_5:=-\gamma_1\gamma_2\gamma_4$.  There are then two different mass terms that can be written in the Lagrangian; viz., ``$m\bar\psi\psi$'' and ``$m\bar \psi \frac{1}{2}[\gamma_3,\gamma_5]\psi$.''  The former is obviously analogous to the natural form in four dimensions and it is invariant under parity transformations.  We therefore define the theory through this term.

In order to study confinement and DCSB we consider the gap equations for the fermion and photon; namely, in a theory with $N_f$ fermions of mass $m$,
\begin{eqnarray}
S(p)^{-1} &=&  i\gamma\cdot p\, A(p)+B(p)\,,\\
&=& [i\gamma\cdot p + M(p)]/Z(p)\,, \\
&=& i\gamma\cdot p + m+\Sigma(p)\,,\\
\Sigma(p) &  = & \displaystyle e^2 \int_q^3 \!D_{\mu\nu}(p-q) \gamma_\mu S(q) \Gamma_\nu(q,p) \,,
\label{gendse}
\end{eqnarray}
where $\int^3_{q} = \int d^3 q/(2\pi)^3$ and, with $q_\pm=q\pm k/2$,
\begin{eqnarray}
D^{-1}_{\mu\nu}(k) &=&\left[\delta_{\mu\nu} - (1-1/\xi)k_\mu k_\nu\right] +\Pi_{\mu\nu}(k),\label{Dprop}\\
\Pi_{\mu\nu}(k) &= &\left[k^2\delta_{\mu\nu} - k_\mu k_\nu\right] \Pi(k) =: T_{\mu\nu}(k)\, k^2\, \Pi(k), \label{Pimn}\\
& = & - N_f \, e^2  \int_q^3 {\rm tr}\, \gamma_\mu S(q_+) \Gamma_\nu(q_+,q_-) S(q_-). \label{PiDSE}
\end{eqnarray}
NB.\, The massless theory is straightforwardly defined by setting the Lagrangian mass $m=0$ and the quenched theory is obtained by writing $\Pi(k)\equiv 0$.

These equations are written in a general covariant gauge, fixed by the parameter $\xi$.  Plainly, they are coupled to each other and to equations that we have not written explicitly; e.g., that for the fermion-photon vertex, $\Gamma_\mu$.  In order to draw reliable conclusions from the analysis of these equations a systematic and nonperturbative truncation scheme should be employed.  One such scheme was introduced in Refs.\,\cite{Munczek:1994zz,Bender:1996bb}.

The coupling $\sigma=e^2$ has mass-dimension one.  Furthermore, QED3 is super-renormalisable and therefore, in contrast to QCD, no ultraviolet divergence can arise whose regularisation would introduce a new mass-scale.  Hence, $\sigma$ sets the scale in the massless theory and there is a direct connection between $\sigma$ and the scales of confinement and DCSB.

With the parametric dependence of the Schwinger functions displayed explicitly, these features entail
\begin{eqnarray}
\label{scaleS}
\sigma\,S(p;m;\sigma;N_f) & = & S(\frac{p}{\sigma};\frac{m}{\sigma};1;N_f),\\
\Pi(p;m;\sigma;N_f) & = & \Pi(\frac{p}{\sigma};\frac{m}{\sigma};1;N_f),\\
\label{scaleGamma}
\Gamma_\mu(p,q;m;e^2;N_f) & = & \Gamma_\mu(\frac{p}{\sigma},\frac{q}{\sigma};\frac{m}{\sigma};1;N_f);
\end{eqnarray}
i.e., that the pointwise evolution of the theory's Schwinger functions at an arbitrary value of $\sigma$ can be obtained through scaling of their $\sigma=1$ behaviour.  In consequence there cannot be a phase transition associated with changes in $\sigma$: whether QED3 is confining and/or exhibits DCSB is independent of the coupling's value.  On the other hand, the state of the theory can respond to changes in the dimensionless parameter $N_f$.

\section{DCSB - Theory}
\label{theory}
In bald terms DCSB is the existence in the chiral limit of a $M(p)\neq 0$ solution to Eq.\,(\ref{gendse}).  Such a solution expresses the generation of mass \emph{from nothing} solely through interactions.\footnote{Whether the vacuum associated with this solution is preferred over that corresponding to the $M(p^2)\equiv 0$ solution; viz., has the higher pressure, can subsequently be determined.  It certainly does in QCD \protect\cite{Bhagwat:2003vw}.}  In order to determine whether that is possible one can begin by analysing the gap equations at leading order in the truncation scheme of Refs.\,\cite{Munczek:1994zz,Bender:1996bb}, which corresponds to using
\begin{equation}
\Gamma_\mu(p,q) = \gamma_\mu\,.
\label{rainbow}
\end{equation}
In this ``rainbow-truncation'' Eqs.\,(\ref{gendse}) and (\ref{PiDSE}) form a closed set, which has long been the subject of scrutiny; see, e.g., Refs.\,\cite{Roberts:1994dr,Bashir:2005wt} and references therein.

\subsection{Landau gauge}
Before reporting the results of such an analysis, we consider it important to discuss whether there is a special choice of gauge parameter for which Eq.\,(\ref{rainbow}) is most sound.  If so, then one can couple that particular gauge choice with Eq.\,(\ref{rainbow}) and define the vertex in any other gauge via the Landau-Khalatnikov-Fradkin (LKF) transform of Eq.\,(\ref{rainbow}) \cite{LK56,Fr56,JZ59}.  This procedure was advocated in Ref.\,\cite{Burden:1991uh} because it guarantees gauge covariance.

Landau gauge, $\xi=0$, occupies this special place.  In Landau gauge the one-loop contribution to $A(p)$ vanishes in any number of dimensions \cite{Davydychev:2000rt}.  Owing to the simplest Ward-Takahashi identity, this feature amplifies the domain whereupon Eq.\,(\ref{rainbow}) can be described as a pointwise accurate approximation.  Additionally, in the context of the reasonable vertex \emph{Ans\"atze} \cite{Burden:1993gy} that have been employed in DSE studies, this means that sensitivity to model-dependent differences between the forms is least noticeable in Landau gauge.  Moreover, whenever it is relevant, $\xi=0$ is a fixed point of the renormalisation group; i.e., the gauge parameter is momentum-independent.  Indeed, it is always zero.

Numerous studies have tried to construct a model vertex that guarantees gauge covariance of the fermion propagator \cite{Bashir:2005wt,Burden:1991uh,Burden:1993gy,Burden:1990mg,Dong:1994jr,%
Bashir:1994az,Bashir:1995qr,Hawes:1996mw,Maris:1996zg,Bashir:2000rv,Bashir:2001vi,%
Bashir:2002dz,Bashir:2002sp,Bashir:2004yt,Fischer:2004nq} and the photon vacuum polarisation \cite{Burden:1991uh,Maris:1996zg,Fischer:2004nq} in QED3 and QED4.  The lesson in the observations above is that such attempts are misguided.  Suppose a Landau gauge \textit{Ansatz} is truly correct, then its form in any other gauge is precisely reproduced by the LKF-transform.  If, on the other hand, the \textit{Ansatz} is wrong, then its in-built dependence on the gauge parameter will not be better than that which is obtained by defining the vertex in different gauges via the LKF transform.  Hence there is nothing to be learnt from exploring the gauge parameter dependence of quantities obtained with a given vertex \textit{Ansatz}.\footnote{In this connection it is materially important to bear in mind that no vertex \emph{Ansatz} can be consistent with perturbation theory if its expression in other than Landau gauge omits an explicit dependence on the gauge parameter; i.e., the correct vertex must possess a dependence on $\xi\neq 0$ in addition to that implicit in the fermion propagator which arises through resolving the Ward-Takahashi identity \protect\cite{Bashir:2001vi}.}  Our view is that irrespective of the choice made for $\Gamma_\mu$, and there are good reasons to employ something more sophisticated than Eq.\,(\ref{rainbow}), the Landau gauge results define the model entirely.

\subsection{Conditions for a critical number of flavours}
In Ref.\,\cite{Appelquist:1988sr}, in conjunction with free massless fermion propagators, $S(p) = -i/\gamma\cdot p\,$, Eq.\,(\ref{rainbow}) was used in Eqs.\,(\ref{Pimn}) and (\ref{PiDSE}) to obtain
\begin{equation}
\label{Pipert}
\Pi(k) = N_f \, \sigma \frac{1}{8 k}\,.
\end{equation}
This result was subsequently inserted into Eq.\,(\ref{gendse}) via Eq.\,(\ref{Dprop}) and, subject to the further truncation $A(p)\equiv 1$,
an equation for $M(p)$ obtained; viz., with $\sigma=1$,
\begin{equation}
M(p) = \frac{1}{2\pi^2 p}\int_0^\infty dq\, \frac{q M(q)}{q^2+M(q)^2} \ln\left[ \frac{p+q+\frac{N_f}{8}}{|p-q|+\frac{N_f}{8}} \right].
\end{equation}
This procedure was described as providing the mass function at leading-order in a $1/N_f$-expansion and yields the result that DCSB is only possible on the domain
\begin{equation}
N_f < N_f^c = \frac{32}{\pi^2} = 3.24\,.
\end{equation}
This claim opened a debate, described in Sec.\,3 of Ref.\,\protect\cite{Roberts:1994dr}.  The other widely canvassed possibility is that the mass function does not vanish but is instead merely exponentially suppressed with increasing $N_f$ \cite{Pennington:1990bx}.  One aim of the following analysis is to eliminate the points of contention.

Given that $\Pi(k)$ must be a monotonically decreasing function in QED3, our first observation is that a critical number of flavours, $N_f^c$, can only exist in the massless theory if in the neighbourhood of this value
\begin{equation}
\label{PiNcr}
\left.\frac{1}{\Pi(k)}\right|_{k\to 0}\stackrel{N_f\simeq N_f^c}{=}0\,.
\end{equation}
To establish this, consider the opposite; namely, $\Pi(0;N_f\simeq N_f^c)= \alpha$, for some $\alpha\in (0,\infty)$.  Under the conditions just described
\begin{equation}
1\geq \frac{1}{1+\Pi(k)} \geq \frac{1}{1+ \alpha}>0 \,,\;\forall k\geq 0.
\end{equation}
Equation~(\ref{gendse}) can only support a gap if the kernel possesses sufficient support.  A lower bound on the support is obtained by employing the replacement: $\Pi(k) \rightarrow  \alpha$.  The equation thus obtained is actually the same as the quenched equation except for a rescaling of the coupling; viz., $\sigma \to \tilde \sigma = \sigma/(1+\alpha)$.  In the quenched theory the coupling serves merely to set a mass-scale and has no effect on whether or not chiral symmetry is dynamically broken.  Thus, if chiral symmetry is not broken for $\Pi(0;N_f\simeq N_f^c)= \alpha$, then there exists a $N_f^\prime<N_f^c$ such that $\Pi(0;N_f^\prime)= \tilde\alpha\neq \alpha$ and chiral symmetry is still unbroken.  Therefore, $N_f^c$ cannot be a critical number of flavours.  Hence no monotonically decreasing function satisfying $\Pi(k)\leq \alpha<\infty$, $\forall k\geq 0$ can generate a critical number of flavours.

If a theory possesses a critical number of flavours, then in the chiral limit and for $N_f\simeq N_f^c$ one can neglect in the Schwinger functions all instances of feedback associated with $B(p)\neq 0$.  Hence Eqs.\,(\ref{gendse}) and (\ref{PiDSE}) can be studied in the forms
\begin{eqnarray}
B(p) &=& \int_q^3\! D_{\mu\nu}(p-q)\frac{B(q)}{q^2 A(q)^2}\frac{1}{4}{\rm tr} \, \gamma_\mu \Gamma_\nu(q,p) , \label{BB0}\\
\nonumber
\rule{-1em}{0ex}p^2 A(p) & = & p^2+ \int_q^3 \!D_{\mu\nu}(p-q)\frac{1}{q^2 A(q)}\\
&  & \times\frac{1}{4}{\rm tr}\, (-i\gamma\cdot p) \gamma_\mu (-i\gamma\cdot q) \Gamma_\nu(q,p),\label{AB0}\\
\nonumber
\Pi(k) & = & -\frac{N_f}{2 k^2}\int_q^3\!\frac{1}{q_+^2A(q_+)q_-^2A(q_-)}\\
&&\times {\rm tr} \left[\gamma_\mu(-i\gamma\cdot q_+) \Gamma_\nu(q_+,q_-)(-i\gamma\cdot q_-) \right] .\label{PiB0}
\end{eqnarray}
(NB.\ Owing to Eqs.\,(\ref{scaleS}) -- (\ref{scaleGamma}), here and hereafter we set $\sigma=e^2=1$.)  In deriving Eqs.\,(\ref{BB0}) -- (\ref{PiB0}) we used the fact that without feedback; i.e., for $B(p) \equiv 0$, all terms in $\Gamma_\mu$ contain a single power of the Dirac matrices.  No term is present that can drive the appearance of another structure.  This entails that the equations for $A(p)$ and $\Pi(k)$ form a closed, coupled pair, one whose solutions will only ultimately influence the result for $B(p)$.

Consider Eqs.\,(\ref{AB0}) and (\ref{PiB0}).  Equation~(\ref{Pipert}) describes the polarisation for $p\gg 1$, on which domain one also has $A(p)-1\approx 0 $.  However, it is the behaviour of these functions for infrared momenta, $p\ll 1$, that is crucial to the existence or otherwise of a gap.  Reference~\cite{Burden:1993gy} indicates the possibility that in the absence of an explicit or dynamically generated fermion mass, $A(p)$ can possess homogeneous power-law behaviour in the infrared whilst nevertheless satisfying its perturbative ultraviolet limit.  We therefore consider the implications of
\begin{equation}
\label{AIR}
A(p) = a_0 \, p^\delta \,,\;\delta > 0\,,\; p\ll 1\,.
\end{equation}

Homogeneity entails $A(\zeta p) = \zeta^\delta A(p)$\,.  From this, the absence of a fermion mass and, importantly, the Ward identity, it follows that
\begin{equation}
\label{GammaIR}
\Gamma_\mu(\zeta p,\zeta q) = \zeta^\delta\,\Gamma_\mu(p,q)\,,\; p,q\ll 1\,.
\end{equation}
Subject to these considerations, inspection of Eq.\,(\ref{PiB0}) reveals that infrared momenta dominate the integration domain therein.  Therefore
\begin{equation}
\label{PiIR}
\Pi(\zeta k) = \zeta^{-(1+\delta)} \, \Pi(k)\,, \; k\ll 1\,.
\end{equation}
NB.\ Dominance of infrared momenta cannot be argued for $\delta<0$.  Moreover, Eq.\,(\ref{PiIR}) is consistent with Eq.\,(\ref{PiNcr}).

Equation~(\ref{AB0}) has a constant term on the left-hand-side.  This is plainly the driving term associated with the ultraviolet behaviour $A(p)=1$, $p\gg 1$, and therefore cannot have any impact on the infrared behaviour of $A(p)$, a result illustrated clearly in Ref.\,\cite{Fischer:2004nq}.  Hence, focusing on momenta within the infrared domain, we substitute Eqs.\,(\ref{AIR}) -- (\ref{PiIR}) into Eq.~(\ref{AB0}) and find
\begin{equation}
\label{AIRc}
A(\zeta p) = \zeta^\delta\,A(p)\,;
\end{equation}
namely an internally consistent result.  The value of $\delta$ depends on the exact form of the dressed-fermion-photon vertex \cite{Fischer:2004nq}, which is not completely constrained by the Ward-Takahashi identity.

It is thus evident that if there exists a critical number of flavours, $N_f^c$, then for $N_f\simeq N_f^c$ the fermion wave function renormalisation function and the photon vacuum polarisation are, for infrared momenta, homogeneous functions.  They have degree $\delta$ and $[-(1+\delta)]$, respectively, a particular relationship that is enforced by the Ward identity.  The converse can also be argued from Eqs.\,(\ref{gendse}) and (\ref{PiDSE}).  If there exists a value of $N_f^c$ such that for $N_f\simeq N_f^c$ $A(p)$ and $\Pi(k)$ are homogeneous functions at infrared momenta, then at this value of $N_f^c$ chiral symmetry is not broken.  This is plain because DCSB would introduce an infrared mass-scale, $\mu$, and the wave function renormalisation and vacuum polarisation would saturate at some constant value for momenta less than this scale.

It follows that there can exist a critical number of flavours for DCSB \emph{iff} the fermion wave function renormalisation function and the photon vacuum polarisation are homogeneous functions at infrared momenta.\footnote{In the fermion gap equation one can represent a Higgs mass, $\eta$, for the photon via an infrared singular polarisation: $\Pi_H(k) = \eta^2/k^2$.  Hence the discussion above also applies in that case.  Reference~\protect\cite{Feng:2005hy} can be viewed as a numerical verification of this.}

These results have an important impact on Eq.\,(\ref{BB0}); namely, for any value of $\delta$,
\begin{equation}
\label{BIR}
B(\zeta p) = B(p)\,,\; p\ll 1\,.
\end{equation}
An \emph{infrared collusion} is apparent in the theory: the infrared anomalous dimensions of the wave function renormalisation and vacuum polarisation cancel from the equation for the scalar self energy and the scaling behaviour of the integrand is $\zeta^0$.  Once again, the Ward identity driven result in Eq.\,(\ref{GammaIR}) is essential for this to occur.  Equation~(\ref{BIR}) states that it is consistent with homogeneity, Eqs.\,(\ref{GammaIR}) -- (\ref{AIRc}), for the scalar part of the fermion self energy to assume a constant value -- possibly nonzero -- at infrared momenta in the neighbourhood of a critical number of flavours, $N_f^c$.

To underline the importance of this outcome, suppose that the fermion wave function renormalisation, fermion-photon vertex and photon vacuum polarisation are treated inconsistently; e.g., such that in Eq.\,(\ref{PiIR}) the exponent is $[-(1+\delta^\prime)]$.  Then the integrand in Eq.\,(\ref{BB0}) scales as $\zeta^{\delta^\prime - \delta}$, in which case the only admissible solution is $B(p)\equiv 0$.  Such a result precludes the existence of a critical number of flavours.  This analysis reiterates and elucidates the observation of Ref.\,\cite{Maris:1996zg}.

\section{Illustration}
\label{sec:illustration}
\subsection{DCSB}
In order to provide a straightforward illustration via the fermion gap equation, Eq.\,(\ref{gendse}), of the results discussed in Sec.\,\ref{theory}, we adapt the vacuum polarisation calculated from numerical solutions of the gap equation in Ref.\,\cite{Burden:1991uh}; namely,
\begin{eqnarray}
\label{CJBPolA}
 \lefteqn{ \Pi(p,q;\ell_{\cal P}) =   {\cal P}(p,q;\ell_{\cal P})\, \Pi(p-q)\,,} \\
\Pi(k) &= & N_f \left[\frac{1}{8} \frac{1}{\sqrt{k^2+ \varsigma a^2}} + \varsigma\,b \,{\rm e}^{- c k^2}\right] .
\label{CJBPol}
\end{eqnarray}
In Eq.\,(\ref{CJBPol}) we use $a=0.20$, $b=0.088$ and $c=7.8$, which owe their nonzero values to DCSB and were calculated \cite{Burden:1991uh} for $N_f=1$.  NB.\ For $a=0=b$, Eq.\,(\ref{CJBPol}) reduces to Eq.\,(\ref{Pipert}); i.e., the leading order result in a $1/N_f$-expansion.

In adapting this form to our purposes we have introduced two new factors.  The quantity
\begin{equation}
\label{sigmarho}
\varsigma = {\rm e}^{-2(N_f-1)/\rho}\,,\; \rho= \frac{M(p=0;N_f)}{M(p=0;N_f=1)}
\end{equation}
allows a suppression of $M(p)$ with increasing $N_f$ to feed back into the vacuum polarisation.  This enables a scenario in which Eq.\,(\ref{PiNcr}) can arise dynamically in concert with the appearance of a critical number of flavours.  We have also included the factor
\begin{equation}
\label{PolP}
{\cal P}(p,q;\ell_{\cal P}) =
\left\{
\begin{array}{ll}
\frac{1}{2} {\displaystyle \left[\frac{1}{A(p)}+\frac{1}{A(q)}\right]}\,, & \ell_{\cal P} = 1\\
1\,, & \ell_{\cal P} = 0
\end{array}\right.\,,
\end{equation}
the presence of which permits the model to realise Eq.\,(\ref{PiIR}) dynamically when $\ell_{\cal P}=1$.  Similar reasoning leads us to write
\begin{equation}
\label{GammaW}
\Gamma_\mu(p,q) = {\cal V}(p,q;\ell_{\cal V}) \, \gamma_\mu \,,
\end{equation}
with
\begin{equation}
\label{PolV}
{\cal V}(p,q;\ell_{\cal V}) =
\left\{
\begin{array}{ll}
\frac{1}{2} \left[A(p)+A(q)\right]\,, & \ell_{\cal V} = 1\\
1\,, & \ell_{\cal V} = 0
\end{array}\right.\,.
\end{equation}
At this point our illustrative model for the gap equation is completely specified.

\begin{figure}[t]
\begin{center}
\includegraphics[clip,width=0.33\textwidth,angle=-90]{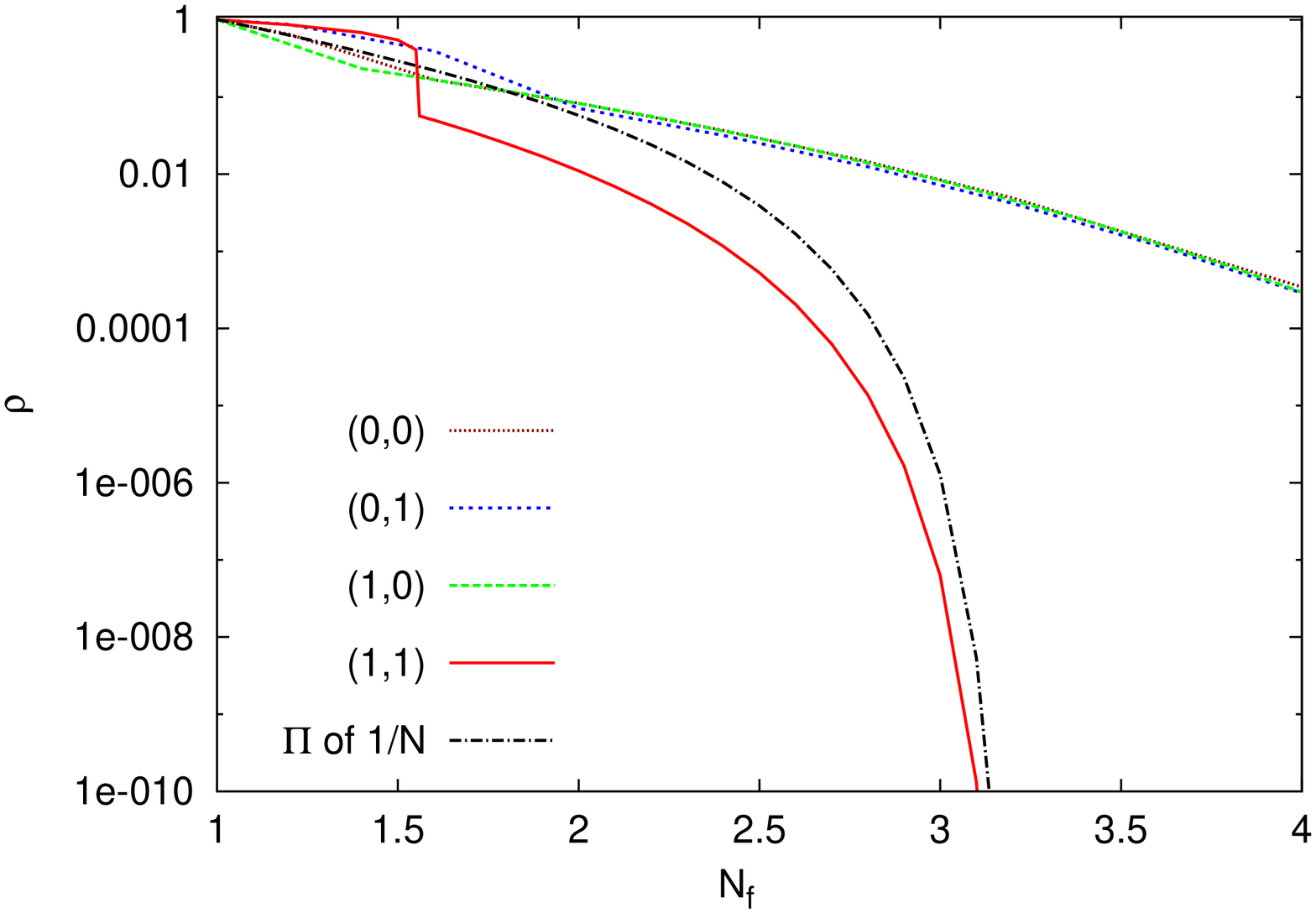}
\includegraphics[clip,width=0.33\textwidth,angle=-90]{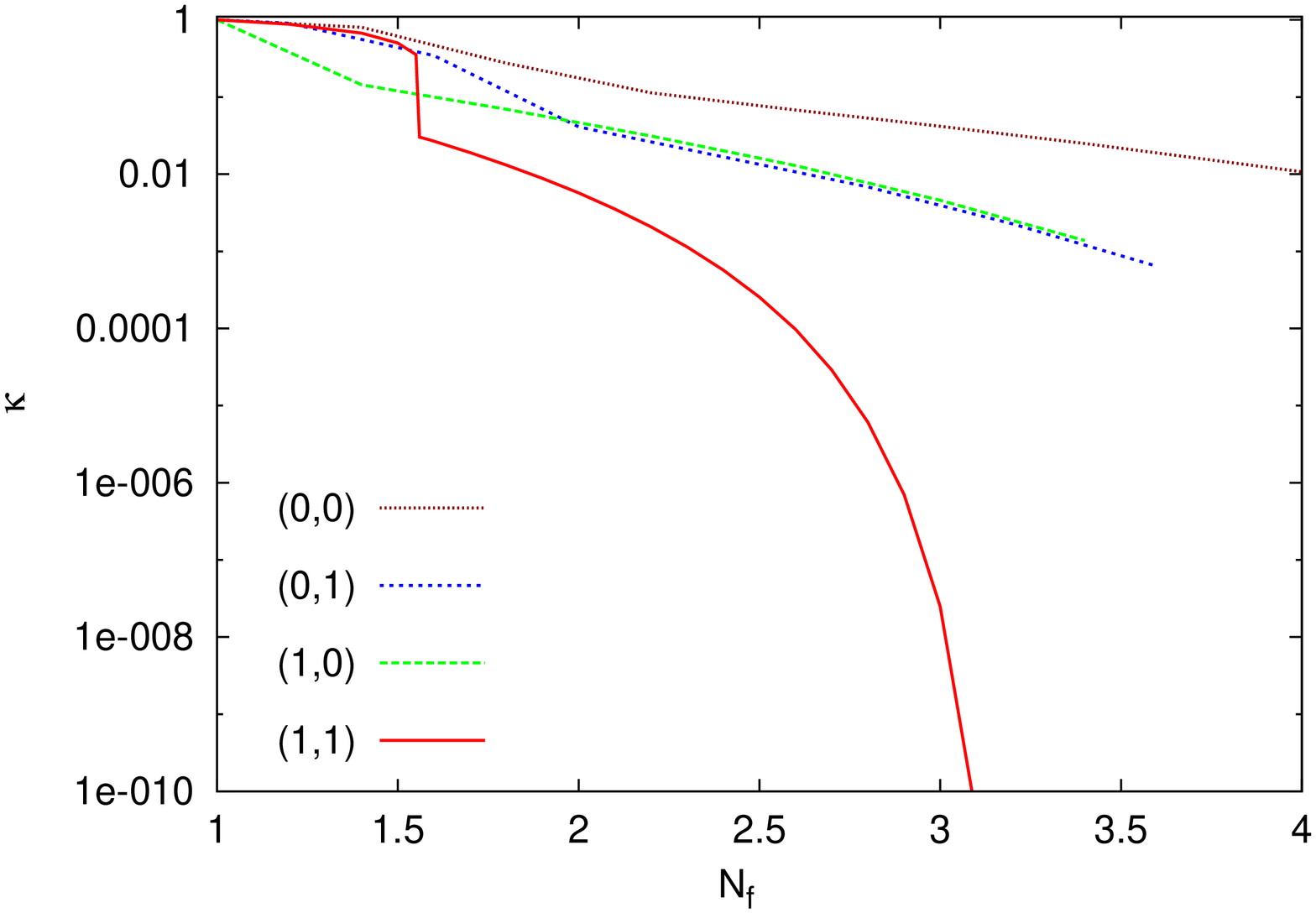}
\end{center}

\caption{\label{Fig1} \emph{Upper panel} -- Calculated evolution of the dimensionless ratio $\rho$ in Eq.\,(\protect\ref{sigmarho}) with the number of flavours, $N_f$: \emph{dotted curve} -- $(\ell_{\cal P}=0,\ell_{\cal V}=0)$ in Eqs.\,(\protect\ref{CJBPol}), (\protect\ref{GammaW}), respectively, which represents the complete suppression of feedback into the gap equation from the solutions of other DSEs; \emph{short-dashed curve} -- $(\ell_{\cal P}=1,\ell_{\cal V}=0)$; \emph{long-dashed curve} -- $(\ell_{\cal P}=0,\ell_{\cal V}=1)$; and \emph{solid curve} -- $(\ell_{\cal P}=1,\ell_{\cal V}=1)$.  Only the last case, which allows a dynamical modification of the photon vacuum polarisation and fermion-photon vertex to influence the gap equation, exhibits a critical number of flavours: $N_f^c=3.24$, Eq.\,(\protect\ref{Nfcdelta}).  \emph{Short-dash--dot curve} -- For comparison, the result obtained via a $1/N_f$-expansion.  \emph{Lower panel} -- The confinement order parameter $\kappa(N_f)$, Eq.\,(\protect\ref{Kconf}).  Simultaneity of chiral symmetry restoration and deconfinement is a robust result.}
\end{figure}

We solve Eq.\,(\ref{gendse}) with $\sigma=e^2=1$, the Landau gauge photon propagator obtained from Eqs.\,(\ref{Dprop}) and (\ref{CJBPol}) -- (\ref{PolP}) and the fermion-photon vertex in Eqs.\,(\ref{GammaW}), (\ref{PolV}).  The upper panel of Fig.\,\ref{Fig1} describes the nature of DCSB in the model.  It is apparent that a critical number of fermion flavours exists if, and only if, the vacuum polarisation and vertex dressing conspire at infrared momenta to completely eliminate the influence of the fermion's vector self-energy from the equation for the scalar self-energy.  Otherwise, $M(p;N_f)$ is merely suppressed exponentially with increasing $N_f$.

\begin{figure}[t]
\centerline{\includegraphics[clip,width=0.33\textwidth,angle=-90]{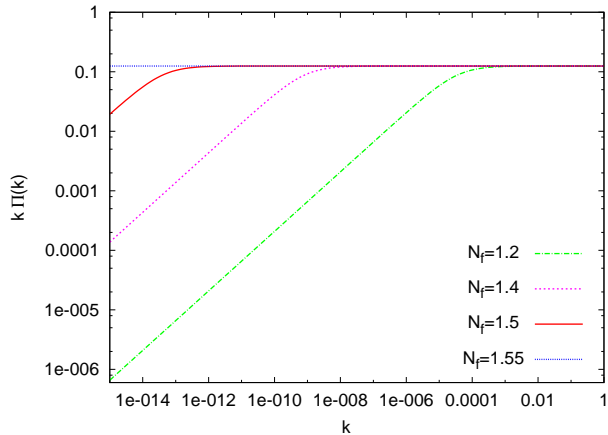}}
\caption{\label{qPol} The polarisation in Eq.\,(\protect\ref{CJBPol}) plotted as $k \, \Pi(k)$ and calculated for a range of values of $N_f$: \emph{dot-dashed}, $N_f=1.2$; \emph{short-dashed}, $N_f=1.4$; \emph{solid}, $N_f=1.5$; and \emph{dotted}, $N_f\geq 1.55$.  The $N_f$-dependence owes to $\varsigma$ in Eq.\,(\protect\ref{sigmarho}) and is self-consistently determined.  Even with the much magnified axes scales in this plot, for $N_f\geq 1.55$ the results are indistinguishable from each other: they lie upon the $1/N_f$-result in Eq.\,(\protect\ref{Pipert}).  }
\end{figure}

The solid curves in Fig.\,\ref{Fig1} exhibit a sudden drop at $N_f=1.55$.  This can be understood via Fig.\,\ref{qPol}, which depicts the vacuum polarisation of Eq.\,(\ref{CJBPol}).  As we advertised, the polarisation acquires a dependence on $N_f$ because a suppression of $M(p)$ feeds back into the polarisation through $\varsigma$.  This mocks up a simultaneous self-consistent solution of the fermion and photon gap equations.  The domain on which $\Pi(k)$ differs from the $1/N_f$-result in Eq.\,(\protect\ref{Pipert}) shrinks with increasing $N_f$, until at $N_f=1.55$, owing to the calculated self-consistent feedback, it no longer contributes measurably to the fermion gap equation.  Thereafter, this gap equation's solution also follows the $1/N_f$-result.

The following observations are relevant.  It is $\Pi(p,q;1)$ in Eq.\,(\protect\ref{CJBPolA}) that self-consistently realises Eq.\,(\protect\ref{PiIR}).  Also, in the upper panel of Fig.\,\ref{Fig1}, which depicts $\rho$, the solid and short-dash-dot curves do not lie upon one another because of the normalisation factor in Eq.\,(\ref{sigmarho}): it is plain from the figure that our \emph{Ans\"atze} yield a $N_f=1$ result for the mass function which is larger than that obtained in a $1/N_f$ expansion.  The curves are indistinguishable if for $N_f\geq 1.55$ we normalise our solution via the $N_f=1$ result obtained using the $1/N_f$ expansion.

It is now clear that in the neighbourhood of chiral symmetry restoration Eq.\,(\ref{CJBPol}) is identical to Eq.\,(\ref{Pipert}), and hence our \emph{Ans\"atze} yield
\begin{eqnarray}
\label{Nfcdelta}
N_f^c = \frac{32}{\pi^2} = 3.24 & \; \mbox{and} \; & \delta(N_f^c) = \frac{1}{12} = 0.083\,.
\end{eqnarray}
We re-emphasise that these values are typical but not definitive.  Extant studies suggest that a more sophisticated treatment of the complex of relevant, coupled DSEs would yield results that differ by $\lesssim 25$\%.

\subsection{Confinement}
Quenched QED3 is confining because it has a nonzero string tension \cite{Gopfert:1981er}.  This feature persists in the unquenched theory but only so long as $\Pi(0)$ is finite \cite{Burden:1991uh}.  As we have seen, $\Pi(0)$ is finite \emph{iff} the fermions are massive: in QED3 only massless fermions can screen completely.

The implications of this for the fermion $2$-point function can be read from Sec.\,2 of Ref.\,\cite{Roberts:2007ji}.  If one writes
\begin{equation}
S(p) = -i \gamma\cdot p \, \sigma_V(p^2) + \sigma_S(p^2)\,,
\end{equation}
then in the absence of confinement ($x=p^2$)
\begin{equation}
\frac{d^2}{dx^2} \, \sigma_V(x) > 0 \,, \; \forall x>0\,.
\end{equation}
On the other hand, $S(p)$ describes a confined excitation
\begin{equation}
\label{RPbroken}
\mbox{if}\;\exists x_c>0:\; \left.\frac{d^2}{dx^2} \, \sigma_V(x)\right|_{x=x_c} \! =0.
\end{equation}
These statements are associated with the realisation of confinement through a violation of the axiom of reflection positivity.  Any $2$-point Schwinger function with an inflexion point at $x=p^2>0$, Eq.\,(\ref{RPbroken}), must breach the axiom of reflection positivity.  This entails that the associated elementary excitation cannot appear in the Hilbert space of observables.

\begin{figure}[t]
\centerline{%
\includegraphics[clip,width=0.3\textwidth,angle=-90]{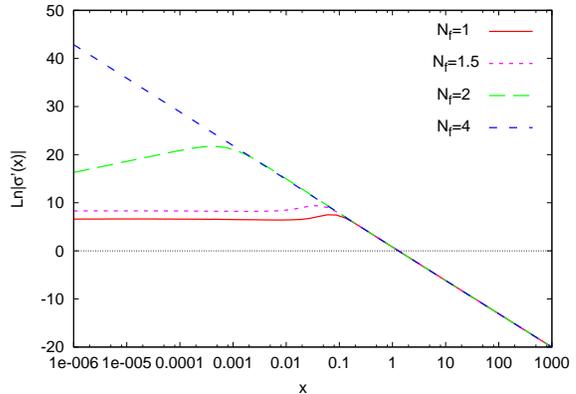}
}
\caption{\label{Fig2} Calculated evolution of $\ln(-\sigma_V^\prime)$: \emph{solid curve} -- calculated with $N_f=1$; \emph{short-dashed curve} -- $N_f=1.5$; \emph{long-dashed curve} -- $N_f=2$; and \emph{dashed curve} -- $N_f=4$.  The conspicuous maximum in $\ln(-\sigma_V^\prime)$ for $N_f=1,1.5,2$ signals fermion confinement in the domain of $N_f$ whereupon chiral symmetry is dynamically broken.  (Recall that $e^2=1$.)}
\end{figure}

In Fig.\,\ref{Fig2} we depict the evolution of $\sigma_V^\prime(x)$ calculated in the model described in Sec.\,\ref{sec:illustration} with three values of $N_f$ below $N_f^c$ and another above.  For $N_f<N_f^c$, chiral symmetry is dynamically broken.  The conspicuous maximum in $(-\sigma_V^\prime)$ for $N_f=1,1.5,2$ signals fermion confinement in the domain of $N_f$ whereupon chiral symmetry is dynamically broken.  The location of the minimum in $\sigma_V^\prime(x)$, $x_c$, migrates to $x=0$ as $N_f\to N_f^{c\,-}$.  One can identify $x_c$ as an order parameter for deconfinement.  NB.\ The location of the minimum positioned farthest from $x=p^2=0$ should be used.  In an asymptotically free theory, if there are any minima at all, then there will be one most distant.

It is evident in the lower panel of Fig.\,\ref{Fig1} that the quantity
\begin{equation}
\label{Kconf}
\kappa(N_f) = \frac{x_c(N_f)}{x_c(N_f=1)}
\end{equation}
exhibits precisely the same behaviour as $\rho(N_f)$.  The figure demonstrates that when internally consistent feedback is enabled, chiral symmetry is not dynamically broken and the fermions are not confined for $N_f\geq N_f^c$.  Plainly, the chiral symmetry restoration and deconfinement transitions are simultaneous in our illustrative model.  The basic causal connection is a dramatic change in the analytic properties of the propagator which accompanies the disappearance of a nonzero fermion scalar self-energy.

It is noteworthy that DSE studies of QCD, in rainbow-ladder truncation with the model kernel of Ref.\,\protect\cite{Frank:1995uk}, find the same results for these transitions at $T\neq 0$ \protect\cite{Bender:1996bm} and $\mu \neq 0$ \protect\cite{Bender:1997jf}.  In addition, Ref.\,\cite{Blaschke:1997bj} describes a model that exhibits a line of simultaneous transitions in the physical quadrant of the $(T,\mu)$-plane.  Indeed, deconfinement and chiral symmetry restoration are coincident in all self-consistent studies of concrete models of continuum QCD that exhibit both phenomena, and in numerical simulations of lattice-regularised QCD \cite{sinclair}.  The methods described herein could be employed to extend the study of QED3 at nonzero-$(T,\mu)$ in
Ref.\,\cite{He:2007jd}, and explore whether and under which conditions chiral symmetry restoration and deconfinement are coincident.

\section{Aper\c{c}u}
\label{apercu}
We argued that QED3 can possess a critical number of flavours, $N_f^c$, above which dynamical chiral symmetry breaking is impossible if, and only if, in the neighbourhood of $N_f^c$ the fermion wave function renormalisation and the photon vacuum polarisation are homogeneous functions at infrared momenta.  The Ward identity guarantees that the fermion-photon vertex also possesses the homogeneity property and ensures a simple relationship between the homogeneity degrees (anomalous dimensions) of each of these functions.  One cannot, however, conclude from this analysis that QED3 must possess a critical number of flavours.  The existence and value of $N_f^c$ are contingent upon the precise form of the fermion-photon vertex.  Notwithstanding this, it is noteworthy that all extant studies which employ a reasonable vertex \emph{Ansatz} and treat QED3's gap equations consistently yield a value $N_f^c\sim 3.5\,$.  It is therefore almost certain that chiral symmetry is dynamically broken for $N_f=2$ QED3, the number relevant in condensed matter applications of the theory.

While our analysis does not establish the existence of $N_f^c$ it does obviate any further debate on the question of whether the fermion mass vanishes at some finite $N_f$ or is merely exponentially damped with increasing $N_f$.  We have elucidated the conditions under which a critical number of flavours can exist and the mechanism by which it then arises; viz., infrared collusion.  All extant studies that report an exponential suppression of $M(0)$ have suppressed that mechanism.

Our discussion should also forestall any further consideration of the gauge dependence of $N_f^c$ and other physical quantities.  We have made plain that Landau gauge occupies a special place in gauge theories.  It is the gauge in which any truncation or sound \textit{Ansatz} for the fermion-photon vertex can most reasonably be described as providing a pointwise accurate approximation.  The vertex in any other gauge should then be defined as the Landau-Khalatnikov-Fradkin transform of the Landau gauge \emph{Ansatz}.  The sensible implementation of this procedure guarantees gauge covariance and hence renders moot any question about the gauge dependence of gauge invariant quantities.

In order to illustrate our arguments we employed a simple model for the photon vacuum polarisation and fermion-photon vertex.  Within this model we calculated $N_f^c$ and the infrared anomalous dimension.  Furthermore, we introduced a new, model-independent order parameter for the confinement of elementary excitations.  In our model deconfinement and chiral symmetry restoration are coincident at $N_f^c$.  This owes to an abrupt change in the analytic properties of the $2$-point fermion Schwinger function when a nonzero scalar self-energy becomes impossible.  It is plausible that this mechanism and result persist in QED3 proper and, indeed, in general.

We remark in closing that all the lessons learnt in this study are relevant to QCD.  The simplest points are that Landau gauge is of particular utility; and confinement and DCSB might in general be tied to, and connected via, the analytic structure of the fermion propagator, which itself is strongly influenced by the possibility of a dynamically generated scalar self-energy.  The phenomenon we have labelled as infrared collusion could be relevant to the phase diagram of QCD in the plane formed by the coupling and the number of quark flavours; e.g., Ref.\,\cite{Gies:2005as,Kurachi:2006mu}.

\begin{acknowledgments}
We are pleased to acknowledge valuable interactions with B.~El-Bennich, T.~Kl\"ahn and R.\,D.~Young.
This work was supported by:
AMC-FUMEC grant and CA23 grant of the Universidad Michoacana de San Nicol\'as de Hidalgo;
CIC and CONACyT grants under projects 4.10, 4.22 and 46614-I;
and
the Department of Energy, Office of Nuclear Physics, contract no.\ DE-AC02-06CH11357.
\vspace*{\fill}
\end{acknowledgments}

\end{document}